**Teachers' Experiences with Implementing Open-ended Labs in High School Physics Classes**

Hamideh Talafian*, Tim Stelzer, Morten Lundsgaard, Maggie Mahmood, and Eric Kuo

University of Illinois Urbana-Champaign

Department of Physics

1110 W Green St

Urbana, IL, 61801

*Corresponding Author. Email: talafian@illinois.edu

## Abstract

Open-ended labs in science can create ample opportunities for high school students to experience authentic science learning. Although most teachers acknowledge the importance of such labs in students' science learning experiences, implementing them in high-school classes can be challenging for teachers. In this work, we conducted a multi-method analysis of data from a teacher Community of Practice (CoP) to investigate (a) barriers to using open-ended labs in physics classes, (b) shifts in teachers' beliefs about the use of open-ended labs in their classes during teachers' engagement in a physics teacher CoP in a partnership program, and (c) a case study of one teacher whose shifts in perceptions about guided inquiry labs led to her successful implementation of a guided inquiry lab in her class. The findings confirm the existence of well-known psychological and structural barriers that can prevent teachers from adopting more open-ended approaches in physics labs. Moreover, we discovered how the interaction of these barriers further complicates the adoption of open-ended approaches in physics classes. The study also revealed a significant gap between teachers' current practices and their desired methods for conducting labs, particularly in terms of structure versus openness. The case study offered deeper insights into how shifts in teaching practices occur through changes in perceptions within a supportive CoP.

***Keywords:*** *Physics Labs, Open-ended Labs, iOLab, Physics Teachers, Professional Development*

## Introduction

Reform-based lab instruction– often synonymous with *investigation-style* or *open-ended* approach, offers students authentic science learning experiences (NRC, 1996; 2012) by allowing them to emulate the work of a scientist (Etkina & Van Heuvelen, 2007; Kozminski et al., 2014; Olson & Riordan, 2012). Specifically, lab experiments with a more open-ended inquiry approach, enable students to make experimental decisions to solve scientific problems, thereby engaging them in the scientific process (NRC, 2000; 2012). Despite teachers' awareness of the potential benefits of this approach in teaching labs, various barriers hinder its implementation. These barriers include time constraints and institutional factors (Dillion, 2008; Settlage & Meadows, 2002; Valli et al., 2008), as well as o teachers' perceptions of their students' abilities (Clermont, Borko, & Krajcik, 1994; Levitt, 2002; Mansour, 2009), teachers' academic backgrounds, and classroom dynamics. Additionally, the lack of access to general science materials (Elechi & Eya, 2015) and lab-specific resources exacerbates the challenges of adopting an open-ended inquiry approach in teaching physics labs.

This paper explores whether the social and material resources available to high school physics teachers in a Community of Practice (CoP) (Lave & Wenger, 1991) can help them overcome the barriers to implementing a more open-ended inquiry approach in teaching labs. Specifically, we seek to address the following research questions:

1. What are the barriers to adopting a more open-ended inquiry approach in teaching labs in high school physics classrooms when the teachers are provided with social and material resources?
2. What evidence, if any, indicates shifts in teachers' perceptions of open-ended lab approaches as a result of their participation in the Community of Practice?
3. Is there evidence of a transition from a more structured lab format to a more open-ended lab approach among high school physics teachers? If so, what factors facilitate this change?

This study contributes to science education research and physics education research in two ways. First, this study sheds light on high school teachers' perspectives on taking a more open-ended approach in teaching labs in science/physics classes, and perceived barriers that may persist, even when equipment and community support are present. Second, this study advances the field toward a potential model for how community support by and for high school science teachers can create transformative and lasting changes in their pedagogical philosophy and classroom lab implementation toward a more open-ended approach.

## Background

### Inquiry-based Instruction, 'Scientific Practices', and Laboratory Work

Over the past several decades, science education reform movements have advocated for investigative approaches that encourage engaging students more in the process of knowledge construction (Miller, Manz, Russ, Stroupe, & Berland, 2018). The early descriptions of this approach have been renamed from *scientific inquiry* to *scientific practices* to shift the focus to what scientists *do* (National Research Council, 2012). Both terms are being used interchangeably in scientific texts. Currently, the framework for K-12 Science Education lists eight scientific practices, which are incorporated into the Next Generation Science Standards (NGSS): (1) Asking questions, (2) Developing and using models, (3) Planning and carrying out investigation, and (4) Analyzing and interpreting data, (5) Using mathematical and computational thinking, (6) Constructing explanations, (7) Engaging in argument form evidence, and (8) Obtaining, evaluating, and communicating information (NRC, 2012).

These scientific practices include a wider domain compared to scientific inquiry allowing students to go beyond experiencing inquiry, by interpreting the results from the evidence and developing arguments, explanations, and models (Crawford, 2014). At the same time, these practices are not happening in a step-by-step fashion; they may interact with each other, iterate by design, or some steps might get eliminated or emphasized more depending on the practice goals (Gericke et al., 2023; Talafian & Ansell, 2023). In the context of physics lab instruction, the American Association of Physics Teachers (AAPT) has enumerated six learning outcome focus areas for physics lab instruction: (1) constructing physics knowledge, (2) developing and testing models, (3) designing experiments, (4) developing technical skills, (5) analyzing and visualizing data, and (6) communicating physics knowledge and experimental results (Kozminski et al., 2014). For instance, in the Investigate Science Learning Environment (ISLE) approach and curricula in physics, these practices have been widely used to encourage students to 'do science' by designing experiments, constructing explanations, and solving practical problems (Etkina, 2007). The ISLE approach in teaching labs has proved successful both in college and high school-level physics classes even in courses that include traditional lectures (Rutberg, Malysheva, & Etkina, 2019; Buggé, 2020; Buggé, 2023). In this partnership program, the teaching approach in labs, the university labs themselves, and the tools provided to teachers all promote the ISLE approach. Our focus here is on fostering a 'doing science' mindset, which refers to a more open-ended approach in physics labs. This approach allows students greater freedom to conduct investigations, from designing experiments to observing data and interpreting results, with minimal guidance.

**Open-ended Labs: Are They Really Effective?**

The focus on engaging students with scientific practices is linked to continued exploration and design of scaffolded, but not heavily prescribed, lab activities. Many educators and science education researchers have found positive evidence in support of investigative science learning approaches such as inquiry-based laboratory work (Abd- El-Khalick, BouJaoude, Duschl, Lederman, Mamlok-Naarman, Hofstein, Niaz et al., 2004; Hodson, 1998; Lunetta, 1998; National Research Council, 2000; 2007; 2012). Empirical evidence demonstrates the significantly positive impact of inquiry-based lab instruction on students (e.g., Chatterjee, Williamson, McCann, & Peck, 2009; van Rens, van der Schee, & Pilot, 2009; Buggé, 2023). For example, Buggé (2023) found that students showed improved scientific abilities in physics when they were given a chance to revise their labs in an ISLE-approach classroom. These students had the chance to brainstorm ideas, observe demonstrations, participate in guided experiments, and engage in meaningful group discussions. Chatterjee et al. (2009) noted that survey results from approximately 700 students working in inquiry-based labs revealed positive attitudes towards these labs, with students believing they learn more naturally and effectively in guided-inquiry settings. Other works have also found that taking a more open-ended approach in teaching labs can support student development of content knowledge (Deters, 2005), be more enjoyable than closed investigations (Lord & Orkwiszewski, 2006), improve students' agility with scientific processes (Etkina & Van Heuvelen, 2007), foster more positive student attitudes toward science (Gibson 1998a, 1998b), and promote student interest in science careers (Gibson & Chase, 2002).

Many labels and categorization schemes have been used to describe the level of prescription and scaffolding in an instructional lab activity. Akuma and Callaghan (2018) considered four levels of inquiry-based instruction: confirmation, structured, directed, and open. They propose that as one moves from confirmation to open instruction, students are more likely to engage in a greater number of the eight NGSS scientific practices. In other categorizations, the levels of inquiry have been placed on a continuum (e.g., Kidman, 2012), more structured labs are referred to as *teacher-driven,* and open-ended inquiry as *learner-driven*. Confirmation style and structured style labs have also been referred to as cook-book style labs where the procedure is given to the learner (Gibson & Chase, 2002). Evidence shows better gains in favor of guided-inquiry lab styles in science labs compared to open-ended inquiry when there is some preparation prior to the instruction or verbal guidance during instruction (Dobber et al., 2017; Furtak, et al., 2012; Gericke at al., 2023; Heindl, 2019). In comparison, studies of "cookbook" style, or more procedurally closed labs do not show significant evidence in support of developing physics content knowledge and expert beliefs about the experimental nature of physics among college-level students (Wieman & Holmes, 2015; Wilcox & Lewandowski, 2018). In such labs, learners conduct routine exercises and rarely reflect on their methodology and findings (Lunetta, Hofstein, & Clough, 2007; Abrahams & Millar, 2008).

In this work, we use the term "open-ended labs" to describe a shift away from a prescriptive, procedural approach. However, the term "open-ended," as commonly used in the literature, does not always accurately capture what Akuma and Callaghan (2018) define as level 4, where students still respond to a question posed by the instructor rather than generating their own. This definition includes open-inquiry and guided-inquiry labs or any design that significantly deviates from rigid, cookbook-style, or confirmation-based labs. Further, we echo the views of those who see lab instruction in physics education as lying along a continuum rather than as a binary choice (Banchi & Bell, 2008; Kidman, 2012; Gericke et al., 2023).

## Barriers to Taking an Open-ended Approach in Teaching Labs

Despite the availability of resources like the ISLE material (Etkina & Van Heuvelen, 2007), and a collective nod from researchers affirming the benefits of investigative-style approaches, teachers are often reluctant to take a more open-ended approach in lab instruction (Cheung, 2007; Dillion, 2008). Some scholars have sought to enumerate and categorize the types of barriers that teachers face in taking a more open-ended approach to lab instruction. Cheung (2007) enumerates 11 barriers including the lack of time, teacher beliefs, lack of effective inquiry materials, pedagogical problems, management problems, large classes, safety issues, fear of abetting student misconceptions, student complaints, assessment issues, and material demands (p.109). Similarly, Ramnarain (2016) found similar barriers and categorized them into extrinsic and intrinsic challenges. Ramnarain considered intrinsic challenges to be related to teachers' competencies, including their perceived understanding of content knowledge, and extrinsic challenges to external factors such as short blocks of time and large class sizes. Taking an instructional design perspective, Akuma and Callaghan (2016) related these challenges to different phases of instruction and categorized them into preparation-phase, implementation-phase, and assessment-phase challenges.

Ultimately, whether a teacher chooses to adopt a more open-ended approach depends on various factors, including the goals of the lab, classroom dynamics, the time of year, the topic at hand, and the teacher's confidence in their disciplinary and technological skills. As a result, they may choose to implement this approach only a few times throughout the school year. As Deters (2005) wisely suggests, the goal is to integrate inquiry-based labs "as often as practical." Even conducting a few inquiry-based labs each year can significantly enhance students' critical thinking, self-confidence, and willingness to engage in scientific inquiry by the time they graduate (p. 1180). This underscores the importance of gradually integrating open-ended approaches into lab instruction, ensuring that students develop a robust understanding and appreciation for "doing science."

**Overcoming Barriers in Communities of Practice (CoP)**

Although there only exist a few reports on the mechanisms by which teachers may overcome structural barriers to implementing open-ended labs, some studies indicate how teachers' perceptions of open-ended lab instruction can become more favorable over time. For instance, sustained and intensive teacher PD focused on inquiry-based instruction (Cheung, 2007) can be an effective strategy for improving teachers' knowledge of inquiry-style experiments and supporting teachers' development of inquiry-oriented identities (Hofstein et al., 2004; Friedrichsen et al., 2006; Lazarowitz & Tamir, 1994; Cheung, 2005). To promote the teaching practices best suited to inquiry-style investigations touted by the NRC, strong arguments have been made for a focus on inquiry-based instruction within pre-service teacher education programs (NRC, 2006; Strat et al, 2022; 2023) and in-service teacher PD (Lederman, 2012). Similarly, in their meta-study of 186 studies on ways teachers can support inquiry-based activities, Dobber et al. (2017) describe targeted teacher training as an effective strategy for overcoming perception-related barriers.

By using a Communities of Practice (CoP) model, we highlight the role of PD context in teacher interactions that lead to shifts in their lab teaching practices. CoP is a situative perspective that explains how context influences social learning in a community whether it be a workplace, a book group, or joining a new family (Lave & Wenger, 1991; Wenger, 1998). The "community" in the CoP refers to a group of people who share a common interest or "shared enterprise" in a particular "domain" or area and engage in *community practices* to learn from each other (Lave & Wenger, 1991). Learning in Wenger's view is a social process, with many of the attributes of an apprenticeship model where less experienced members learn from more experienced ones. In the process of becoming a member in the CoP- termed legitimate peripheral participation- one may start from being a "peripheral member" and then gradually transition to a "core member" through socialization, observation, and engagement over time, or remain a peripheral participant (Lave & Wenger, 1991; Wenger, 1998). We believe that in this process, there are lots of exchanges of content and pedagogical and technological knowledge which in our case happens in a professional development setting context. Opportunities that can transform teachers' beliefs in one or more ways that Bandura (1997) names: (1) experiencing success by themselves; (2) observing success by others (3) emotional arousal, and (4) verbal persuasion - all of which can happen in PD settings. Therefore, it is important to investigate teachers' self-reported barriers in the presence of the community and trace gradual perception and practice shifts within the community.

**Context of the Study**

In this paper, we investigate teachers' implementation of open-ended labs in the *Illinois Physics and Secondary Schools (IPaSS)* partnership program and the prevalence of implementation barriers among IPaSS teachers. This partnership is between the University of Illinois Urbana-Champaign (U of I) and Illinois high school physics teachers (Teaching Fellows).

In the IPaSS program, teachers engage in prolonged physics-specific professional development (PD) with 100 hours of in-person and online professional development for up to four years. In this partnership, we aim to create a professional community of physics teachers by (a) sharing research-based, university-level physics materials; (b) facilitating teacher sharing of course materials with one another; (c) supporting teachers in implementing new course materials and activities throughout the year; and (d) supporting teachers in eventually becoming leaders and mentors in the program. Currently, the program is in its fifth year and the data collection and analysis were conducted during years 3 and 4 of the program. Over time, teachers and the IPaSS team have formed a Community of Practice (CoP) where teaching materials, teaching experiences, and support are shared.

       In IPaSS CoP, teachers engage in professional development (PD), with a focus on curricular integration of the iOLab (Figure 2), a multi-sensor lab device that is used in teaching the introductory physics labs of the University of Illinois. The iOLab can be easily deployed to conduct hundreds of physics labs without requiring teachers to use any other lab equipment. Although the iOLab was created with university students in mind, it has been piloted successfully in secondary school classes (Cunnings, 2016). The iOLab software allows teachers and students to analyze graphical data and measure a range of quantities (see Figure 3). Teaching Fellows and their students have free access to a class set of these devices as part of the program. According to Selen & Stelzer (2013), the iOLab is effective in promoting students' freedom in designing their own lab experiments. It is worth noting, that the university introductory labs are open-ended and inspired by the Investigative Science Learning Environment (ISLE) approach (Etkina & Van Heuvelen, 2007) in teaching labs where the focus is on empowering students to get creative in designing solutions to real-life situations.

During the PD meetings, teachers are introduced to the iOLab by conducting some of the university labs and are allowed to adapt these labs to their classroom and/or develop their own iOLab-based labs. However, it is not a requirement of the PD, that teachers choose an open-ended style for their labs or use the university labs. Throughout the school year, teachers have access to direct support from the developers of the lab device, and high school classroom-specific support from peers attempting similar lab reforms. By providing these materials and support opportunities within the CoP, we aimed to give teachers the tools to overcome some of the barriers that they might have experienced otherwise.

**Figure 2**
*iOLab System*

                                                 **Figure 3**
                                           *An example of a kinematic lab with iOLab and data from the wheel sensor.*

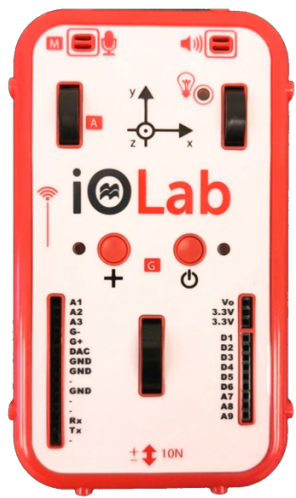

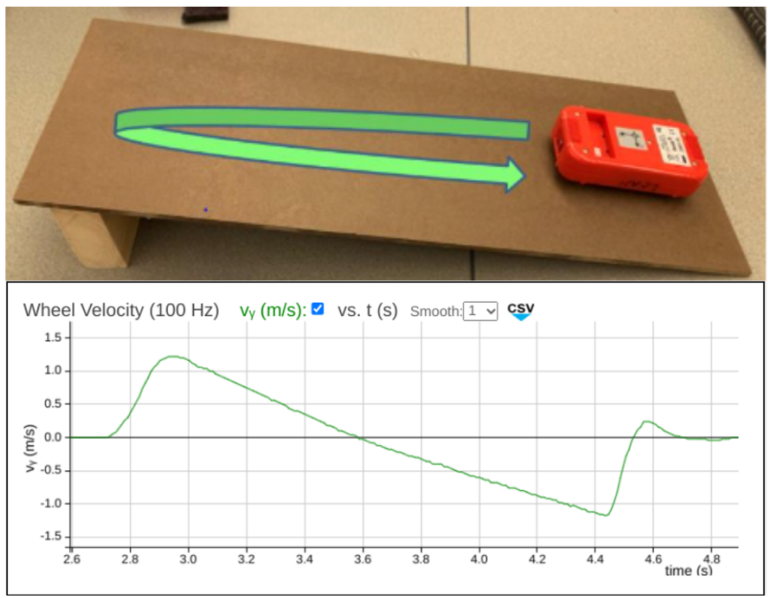

## Methods

### Participants

Teachers in the IPaSS program come from diverse academic and teaching backgrounds some of them holding degrees outside of physics teaching a variety of physics courses spanning from general physics to Advanced Placement (AP) Physics courses. The range of experience is from 0-32 years, and more than half of them teach at Title 1 schools. Using a convenient sampling method, the participants in this study- 34 teachers- are a subset of a bigger population (40 teachers) who are currently enrolled in the program. Table 1 shows the participating information of teachers.

**Table 1**

*Participating Teachers' Information. Total number of teachers are 34.*

| Category | Details | Count |
|---|---|---|
| Locale | Urban | 7 |
| | Suburban | 17 |
| | Town | 6 |
| | Rural | 4 |
| Degree | Physics | 14 |
| | Non-physics | 20 |
| Years of experience | 0-5 years | 10 |
| | 6-10 | 9 |
| | 11-15 | 3 |
| | 16-20 | 6 |
| | 21-25 | 3 |
| | 25 + | 3 |

| Courses taught* | AP Physics | 25 |
| | Regular physics | 29 |
| | Other sciences | 13 |
| Title 1 or > 40% Low Income | Yes | 18 |
| | No | 12 |
| | No data available | 4 |

*The number of teachers overlaps.

## Data Sources

To answer the research questions, we collected various data in each phase for triangulation purposes (Creswell, 2012) (See Table 2). Aligned with RQ1, we collected text, video, and pictorial data to qualitatively inquire about teachers' self-reported barriers to implementing open-ended labs. To start, we recorded a subset of five teachers' reflections on their attempts at implementing open-ended labs in their classes, as part of an online one-hour PD. The online PD meeting happened during Fall 2022 (Dec), and the last 18 minutes, of the video when teachers discussed their views about different types of labs in their classes, was transcribed. The teachers were prompted to talk about assessing labs, lab structures (structured vs. open-ended), and scaffolds in physics labs. As a follow-up to this conversation, we created two open-response survey questions to document 22 teachers' views (13 women, 9 men) on lab instruction. The open-response survey items came from a longer survey that was administered in the Spring 2023 semester (February). The survey asked teachers to reflect on the value of the meetings, and recommendations for improvements, and two questions were added to capture a wider range of teachers' views about lab instruction (including barriers to, and value of, implementation). Question (1): *What do you consider to be some of the challenges with running open-ended / less structured labs? If you have tried more open-ended labs, how has this gone for you and your students?* Question (2): *What do you value in terms of student skills built from labs? How are students doing with these skills this year in general?*

Aligned with RQ 2, we took a quantitative approach in considering teachers' past and present perceptions of open-ended labs, and their future ideal use of this instructional style. To pursue this aim, we collected retrospective survey data from 14 teachers, 7 women, and 7 men, who had been in the program between one and three years to understand how they believed their current perceptions of open-ended lab instruction compared to that when they first joined the program between one and three years ago. Here, we adapted the Guided Inquiry Survey (GIS) from Cheung (2011) in Chemistry open-ended inquiry labs to probe physics teacher perceptions of open-ended labs across three constructs: (1) the value of guided inquiry labs, (2) the limitations of cook-book style labs, and (3) implementation issues of guided inquiry labs. We deployed a retrospective survey in June 2023 to better capture the effect of the community on perceived shifts in their beliefs about open-ended labs. Retrospective surveys have participants reflect on past events at a single point in time (Moore & Tananis, 2009; Pratt, Mcguigan, & Katzev, 2000). Examples of survey questions in this phase, answered on a 5-point Likert scale were: "Most students like guided inquiry experiments more than structured inquiry experiments" and "Guided inquiry experiments can provide more opportunities for students to apply physics knowledge than structured inquiry experiments." To see a full list of GIS questions and constructs see Appendix A.

The embodied rating task was a data source that was used to inform both research questions 1 and 2. In this task, a total of 26 teachers participated, 18 men and 8 women. The ranking task occurred during the in-person PD in August 2023 with 31 teachers who had been in the program between 0-4 years. In this task, the teachers were asked to physically stand somewhere between the continuum of structured vs. open-ended labs based on how they are implementing labs in their classes. Next, they were asked to reposition themselves if they desired to do labs differently. During this task, the PD facilitators prompted them to talk about their approach to lab instruction and why they repositioned themselves.

Addressing RQ 3, the data sources focused on developing a case study of Dawn, a novice physics teacher with a biology background in a small rural school, who teaches biology, chemistry, zoology, and astronomy. Dawn had been in the program for two years and had shown a shift in her practices toward taking a more open-ended approach in teaching labs. For this case study, we conducted two hours of observation from two classes in Dawn's classroom and conducted a 30-minute semi-structured interview with Dawn during the Spring of 2023 (May). Additionally, we analyzed a 15-minute video from an online one-on-one meeting between Dawn and one of the PD facilitators regarding her experience of using an open-ended lab in her class for the first time. Later, she did a workshop in the Summer of 2023 (August) and shared her experience with other teachers. All data was collected within nine months- between December 2022 to August 2023.

**Table 2**

*Alignment of Research Questions with Data Sources.*

| Research Questions | Data Sources | Analysis | Type of Data |
|---|---|---|---|
| **RQ1:** What are the barriers to adopting a more open-ended inquiry approach in teaching labs in high school physics classrooms when the teachers are provided with social and material resources? | • One hour of online PD video (subset of 5 teachers)<br>• Open-ended Survey (21 teachers)<br>• rating task (26 teachers) | • Started with deductive coding based on a priori codes from literature<br>• Continued with inductive approach for emergent codes | • Text data<br>• Video data<br>• Pictorial data |
| **RQ2:** What evidence, if any, indicates shifts in teachers' perceptions of open-ended lab approaches as a result of their participation in the Community of Practice? | • Guided Inquiry Survey (GIS) (14 teachers)<br>• Embodied rating task (26 teachers) | • Descriptive statistics<br>• Shapiro-Wilk test of normality<br>• Paired-sample t-test | • Numerical Likert scale data 1 - 5 (Strongly Agree-Strongly Disagree)<br>• Pictorial data |

| **RQ3:** Is there evidence of a transition from a more structured lab format to a more open-ended lab approach among high school physics teachers? If so, what factors facilitate this change? | • Classroom observation (45 minutes/one period) • 30-minutes interview • 15- minute online planning video | Qualitative inductive coding | • Text data • Video data |
|---|---|---|---|

## Data Analysis Procedure

The PD videos and the case study interview were hand-transcribed and together with the open-ended survey results were coded inductively with MAXQDA software using grounded theory techniques (Strauss & Corbin, 1998). Other qualitative data such as classroom observation and planning videos were not fully transcribed, but some excerpts were transcribed to corroborate with other forms of data. For RQ2, the GIS survey items for each construct were added up, averaged, and plotted. For this survey, we used a paired sample t-test to compare the self-reported perspectives of teachers before joining the program (pre) and now (post). It is worth noting that in the retrospective GIS survey, there was no time gap between the collection of pre and post. The participants were asked to reflect on their views before joining the program and after it, but at a single point in time.

We conducted a pictorial analysis of the embodied rating task by creating and digitizing two images that represent the two parts of the task. By taking screenshots from the workshop videos, we were able to accurately position teachers along a continuum, replicating where they had stood. We then created a second image showing where they moved to when they adjusted their positions. Through these images, we created arrows that indicate their initial and final positions, as well as the direction and extent of change along the continuum.

The qualitative and quantitative analyses were conducted by one researcher (HT). For RQ1, a priori coding scheme based on a review of the literature was used to code 142 segments (111 survey segments and 31 online PD transcript segments), and then the emerging codes were added to the scheme. After initial coding, the final list of nine codes was discussed between authors and three categories emerged. Nine months after the initial coding, the same coder re-coded the data to test the intra-rater reliability of the coding scheme. Four instances of partial disagreement led to adding three new codes in the coding scheme (indicated in Table 2): Students' lack of interest, lack of focus, and inaccessibility to materials. The intra-rater Cohen's Kappa reliability for each of the 12 codes was 0.93. For Phase 3 of the study, an inductive approach was taken to record one teacher's experience with implementing open-ended labs in her physics class. In her interview and subsequent coding, the focus was on extracting the positive and negative experiences that she had with this approach.

**Findings**

**RQ1: Teachers' Self-reported Barriers in Taking an Open-ended Approach in Teaching Physics Labs**

The result of our analysis for RQ1 -teachers' self-reported barriers- showed a genuine interest from teachers to take a more open-ended approach in teaching labs, but they felt hindered by several constraints. In an online PD, an open-ended survey, and a ranking task, three challenges in implementing an open-ended approach in teaching labs emerged: time constraints, teachers' perceptions of their student's abilities, and teachers' perceptions of their own abilities. Table 2 summarizes all codes and categories used in the analysis. Below, we go through each of these three findings in more detail.

**Table 2**

*Coded Segments of Barriers to Implementing Open-ended Labs in Physics Classes*

| Category | Code | Teachers Mentioned the code at least once | Data Source |
|---|---|---|---|
| Structural Barrier | Time in class | Lisa, Arnav, Patrick, Daniel, Sabrina, Grant, Veronica, Marcus, Emily, Philip, Dawn* | Survey and online PD |
| | More work for teacher | Henry | Survey |
| | Populous class | Susan, Henry | Survey |
| | Difficulty in assessment | Patrick*, Emily* | Online PD |
| | Access to materials** | Marcus | Survey |
| Teachers' perceptions of **students'** abilities | Students' lack of interest in investigation** | Amy | Survey |
| | Students' lack of focus in investigation** | Marcus | Survey |
| | Students' lack of familiarity with investigative teaching | Patrick, Jeff, Serena, Tony, Philip, Katie | Survey |
| | Students' lack of confidence | Francesca, Kayla | Survey |
| | Unproductive struggle | Daniel, Sophia, Henry, Patrick* Arnav* | Survey and online PD |
| Teachers' perceptions of their **own** abilities | Teachers' lack of content knowledge | Emily, Dawn | Survey and online PD |
| | Teachers' lack of technological knowledge | Veronica, Dawn* | Survey and online PD |

*Mentioned the code only in online PD.

**New codes added to the coding scheme after 9 months of initial coding.

***Structural Barriers***

Analyzing teachers' statements revealed some structural barriers (Anderson, 2007; Deters, 2004) teachers had experienced, or considered, in implementing open-ended labs. The most frequently cited structural barriers were limited class time, increased workload for teachers, large class sizes, challenges in assessment, and limited access to materials. *Time* was cited most frequently (11 teachers out of 21) as a barrier that hinders teachers' desire to take a more open-ended approach. Depending on the school, teachers in our program had periods as short as 40 minutes to blocks as long as 90 minutes. For teachers with shorter blocks, time management was a bigger challenge when it came to taking a more open-ended approach to labs.

It is notable that 11 teachers, named *time* available in the class as the biggest obstacle. Many cited how time constraints can become exacerbated by other structural barriers such as large class sizes and short blocks of planning time. For example, some teachers described how time constraints, combined with larger class sizes, can catalyze an even more stressful teaching scenario (See Table 3 for stacking barriers). Two veteran teachers (>25 years of physics teaching experience) with large class sizes (>24) describe this stacking of structural barriers. Susan focuses on class size as the main barrier, hinting that individual attention and the ability to monitor students' work are key in preventing small flaws in procedural design or data collection that can snowball into a confusing result during the analysis phase. Henry, a teacher in a medium-sized, Title 1 rural setting school references the challenges of a populous class that create more work for teachers in taking a more open-ended approach, *"[having a more open lab prompt] required work on my end to meet with every group and make sure they aren't going down a path that wastes their time, and it is harder now that my class sizes are larger this year"* which was coded both for *time* and *more work for teachers* based on the coding scheme.

### Teachers' Perceptions of Their Students' Abilities or Feelings

Fourteen teachers (66.6%) cited students' abilities in dealing with investigative-style challenges as a barrier to implementing open-ended labs. Hence, teachers, advocate for more guided and structured labs with prescriptive procedures to support students in reaching scientific conclusions. Survey data also revealed that teachers' low perceptions of their students' abilities were related to (1) students' unfamiliarity with investigative science learning approaches (6 teachers), (2) students' "unproductive struggle" when completing labs with fewer scaffolds (5 teachers), (3) students' lack of interest, focus, or confidence, which leads to anxiety and giving up on lab tasks (4 teachers) (See Table 2). However, at the same time, they express a desire to train students to tackle the challenges of investigative approaches if the *time* barrier is absent.

Examples of those who believed students' lack of familiarity with investigative science learning are causing the challenge include: *"Students are not often asked to be creative and have difficulties setting labs up from scratch"*, or *"I think it's very hard for students who are not used to this to adjust to."* These statements describe a type of student discomfort that appears to stem from the lack of prior exposure to more open-style tasks in their education. Similarly, Katie writes, *"The lower-level students haven't gained the inquiry skills to be "set free" just yet. I believe it could still be a product of the COVID years [...] but students struggle immensely with answering open-ended questions, let alone designing an open-ended lab."*

Among teachers who referred to the students' struggles in handling more open-ended style labs, we have Tony, a novice physics teacher who, despite his genuine interest in open-ended labs, believes his students *"do not even know where to begin."* The same concern was also raised by other teachers like Patrick and Arnav. Patrick believed students' struggles become *"distracting"* and will eventually eat up so much of class time. Arnav raised this concern by giving an example during the online PD where Patrick and Dawn were also nodding as a sign of agreement *"the ramp acceleration that we just did, I put a big highlighter, please check this with [his name] … because students don't understand when they let go the cart or when it's at the peak and I had to question each group, so tell me where is it at the peak, and they point to the wrong part of the graph."* To help students get unstuck, Henry thinks teachers end up doing more work "*It required work on my end to meet with every group and make sure they aren't going down a path that wastes their time, and it is harder now that my class sizes are larger this year. I also find that if I don't make students call me over when they are analyzing the IOLab graphs, they will misinterpret the data, at least at the early part of the semester.*" However, like most of the teachers in the program, he still sees benefits in incorporating such labs, at least in theory: *"Another reason I like open-ended labs is the students sharing what they've learned at the end, and theoretically there's more learning in the class if the students are all answering slightly different questions.*

Teachers also reflected on students' feelings during more open-ended labs, reporting feelings of student frustration or anxiousness when they had tried an open-ended lab task. For instance, Kayla and Tony found one motivating factor in their use of more prescriptive labs is that this style can offer students a solution to a problem after just 45-60 minutes. This reduces the possibilities of student frustration and provides students with a more satisfying sense of closure and success each class period. Amy recognizes that open-ended labs require significant scaffolding, but also that there is a point at which the tasks become increasingly closed when scaffolding begins to pass into a territory that feels more like handholding. Amy expresses, *"When I try to take away some of the scaffolding, the kids in 3rd and 7th period just sit there and stare at each other. They have no interest in exploring, they just want a recipe to follow."* Kayla also mentions: *"Open-ended or less structured labs can be a challenge because students tend to be not confident in the material, which leads to students more frequently giving up."* Patrick works in a small, rural setting Title 1 school, and faces similar challenges. In the online PD session, he described concern about making his lessons too challenging and "*scaring students off*", referencing a culture of avoiding academic challenges in his school community. In the survey, Patrick cites concerns about multi-day investigations that may not yield sufficiently impactful conclusions to justify the time invested. Ariel, with a small class roster like Patrick, considers scaffolding as a silver bullet that can reduce student anxiety while navigating looser lab designs. She writes, "*I find that most students become anxious when they don't have step-by-step procedures on a lab handout. From my experience, this highlights the work teachers must do to help students become comfortable with productive struggle.*" She references *"productive struggle"*, noting that it is a teacher's job to *"help students become comfortable"* by exposing them to these more open tasks. It is worth noting that Ariel has a class of under 10 students in the year of the survey and may feel more confident in her ability to provide effective real-time scaffolds to her students as a result.

### Teachers Perceptions of Their Own Abilities

The third category of barriers based on our analysis is teachers' perceptions of their own abilities, which refers to both their content knowledge (Physics) and technological knowledge (iOLab proficiency). This concern was most strongly articulated by two novice teachers holding non-physics degrees. During the online PD, these two teachers saw their insufficient physics content knowledge as a barrier to implementing open-ended labs. They were mainly worried about the content-related questions coming from students during lab work that do not follow a rote procedure that they can pre-rehearse. They referred to their potential inability or uncertainty in quickly and correctly answering questions that may spontaneously arise in an open-ended lab setting as a significant barrier: *"Part of my struggle is that I lack the background knowledge to steer the students in the right direction without giving too much guidance. It also takes so much time."*

Similar to teachers' concerns about their inability to provide immediate feedback to students' content-related questions described above, here they expressed concerns about technology-related questions that might come up in open-ended labs. Teachers cited the possibility of students choosing from multiple iOLab sensors to complete the same open-ended activity, which would require teachers to be proficient with multiple sensors and methods that may come up. While teachers appreciated having access to the U of I team, they still found the delay of up to 24 hours in getting a response too long for their students *"[students] need the right answer right away and they can't wait,"* Dawn, a novice teacher added.

### Multiple Barriers at Work

In the survey, teachers commented on taking an open-ended approach in a way that revealed the interwoven nature of multiple structural and/or perception-related barriers in teaching labs. Table 3 shows some of these segments for which multiple codes were assigned along with "Time." Some table entries describe how other structural barriers can stack to make the time barrier feel more significant (e.g., Susan), whereas others describe how their perceptions of students' abilities, or their own competency can make them more wary of running out of time for an investigation in the class (e.g., Patrick, Lisa, Emily, Philip).

Table 3 shows the three most common stacking of barriers. For example, the above-cited Henry's concern over one's ability to be vigilant of all lab groups' procedures *could* be read as a structural constraint if it is tied to class size, but, underlying this concern is also a *perception* of student abilities: that students are likely to go down an unproductive route, or accidentally cement the learning of incorrect physics concepts, if consistent real-time guidance and feedback is not provided. Whereas most teachers who cited time constraints were concerned that students would fritter away class time by getting stuck on developing their own procedure, some teachers (e.g., Emily) held the belief that they would slow the students down due to not being able to adequately answer their technological or physics content questions.

**Table 3**

*Some Coded Segments that Presented Multiple Barriers*

| Teacher name (pseudonym) | Segment | Assigned Codes | Category |
|---|---|---|---|
| Susan | *"Large class sizes and the need for smaller group sizes to maximize student participation make it challenging to help individuals."* | Time<br><br><br>Populous classroom | Structural barriers<br><br>Structural barriers |
| Lisa | *"Giving students the time needed to play and try things out is one of the hardest things. We just don't have the large blocks of time that a college class has weekly. There's no way to allow for 3 hours of lab time per week; sometimes 1 hour is tough to fit in. I also found on a recent lab that a number of groups were using what I would consider to be an invalid set of procedures, but I missed this because I was pushing for them to design their own rather do it my way."* | Time<br><br><br>Unproductive Struggle | Structural barriers<br><br>Teachers' perceptions of students' abilities |
| Emily | *"Part of my struggle is that I lack the background knowledge to steer the students in the right direction without giving too much guidance. It also takes so much time."* | Time<br><br>Teachers' lack of physics knowledge | Structural barriers<br><br>Teachers' perceptions of their own content and technological abilities |

Our analysis also revealed that central to the selection of the lab styles was teachers' lab goals which showed to be impacting teachers' decisions. For instance, during the online PD, one teacher talked about the significance of lab goals as a factor that can determine how teachers pick a particular style of lab over another. Other teachers in attendance nodded in agreement when Emily said *"A lot of [teacher decisions] come to the ultimate goal of the lab. Sometimes my goal is just to take data and apply the data in specific ways. But sometimes the goal is to assess a specific problem. I think all types of labs have a place in science. Sometimes you need them to analyze something to eventually be able to do an open-ended lab."* Moreover, some teachers, like Patrick, placed a philosophical emphasis on concluding a topic or coming to a full group conclusion aligned to a physics-specific learning goal within a single class period. Even experienced teachers with a strong drive to promote student creativity and critical thinking feel conflicted about how to strike a balance when confronted with time constraints. For instance, Lisa writes about wanting to allow students to *"play"* but also finds this challenging in a single 1-hour period.

Yet these teachers see these barriers as surmountable with the right strategy or a shift in mindset about the goals they have for their students in their own context. As part of the sharing in the PD, some teachers embrace the messiness of open-ended labs in the high school setting as a learning process. For example, Philip views the goal of early iOLab experiments as teaching students to become comfortable tinkering in the lab, as opposed to teaching them neatly packaged physics concepts that fit within canonical textbook physics. Yet, at the same time, he articulates the process of student learning as a long game that often benefits subsequent years.

## RQ2: Shifting in Teachers' Perceptions of Implementing Open-ended Labs in Physics Teaching Communities of Practice (CoP).

For RQ2, we explored teachers' perspectives on implementing open-ended labs using the Guided Inquiry Scale (Cheung, 2011) survey, and an in-person ranking task. The Guided Inquiry Scale (GIS) survey was administered at the beginning of the fourth IPaSS summer PD. At this point, cohorts 1-3 teachers had been in the program for at least one year, so we were able to gain insights into their perceived changes in attitude toward open-ended labs as a result of participating in the program. To this end, we administered the survey in a retrospective form (Moore & Tananis, 2009; Pratt & Mcguigan, 2000) and asked them to answer each item by comparing their attitudes from before joining the program to the moment of completing the survey. The results are presented in Table 4.

**Survey.** The GIS survey measures three constructs: the value of guided inquiry labs, the limitations of cook-book style labs, and implementation issues of guided inquiry labs. Paired sample t-test results reveal a difference in teachers' *value of guided inquiry labs*, *limitations of cook-book style labs,* and *implementation issues of guided inquiry labs* during their engagement with the program for these physics teachers at a p < .001 level. As shown in Table 4, teachers valued open-ended inquiry labs more after participating in the program. Additionally, there was an increased awareness of the limitations of cookbook-style labs post-program. Finally, the results also revealed improved perceptions related to implementing guided inquiry labs. It is worth noting that the GIS survey constructs are differentiating between guided-inquiry style labs and cook-book style labs, and we are not equating guided inquiry with open-inquiry. What is important here is that we are investigating teachers' preferences in terms of shifts in taking a more open-ended approach which can have different levels of scaffolds as opposed to implementing open-ended style labs.

**Table 4**

*Paired Sample T-test Results of GIS Survey with 14 Teachers*

| | Mean and Standard Deviation | | | | |
| --- | --- | --- | --- | --- | --- |
| | Pre-test (Before joining IPaSS) | Post-test (After joining IPaSS) | *t* | *p* | *d* |
| Value of Guided Inquiry | 3.43 (0.92) | 4.52 (0.37) | 4.84 | <.001 | 1.3 |

| | | | | |
|---|---|---|---|---|
| Limitation of Cook-book Style Labs | 3.32 (0.82) | 4.20 (0.72) | 4.83 | <.001 | 1.3 |
| Implementation Issues of guided-inquiry Labs | 3.14 (0.77) | 3.84 (0.55) | 3.55 | .004 | 0.95 |

**Rating Task.** After discovering that teachers perceived significantly greater value in open-ended labs after one or more years in the program, yet still faced substantial obstacles to implementation, we aimed to explore how these perceptions of implementation challenges influenced their future practice goals. To this end, an in-person embodied rating task was administered shortly after the GIS survey and included teachers who had joined at any point in the first four years of IPaSS. The embodied rating task took place at the fourth summer PD and gave us a qualitative sense of how far teachers felt they were from meeting their future goals concerning open-ended lab implementation. The task required teachers to line up twice. First, they were asked to position themselves according to where they believed they currently were in terms of teaching physics labs, with the left side of the room representing open-ended labs and the right side representing structured labs. The second time, they were asked to stand where they desired to be on this spectrum when teaching labs in their class.

**Figure 5**

*Teachers' Movement Between Their Current and Desired Positions in Teaching Physics Labs*

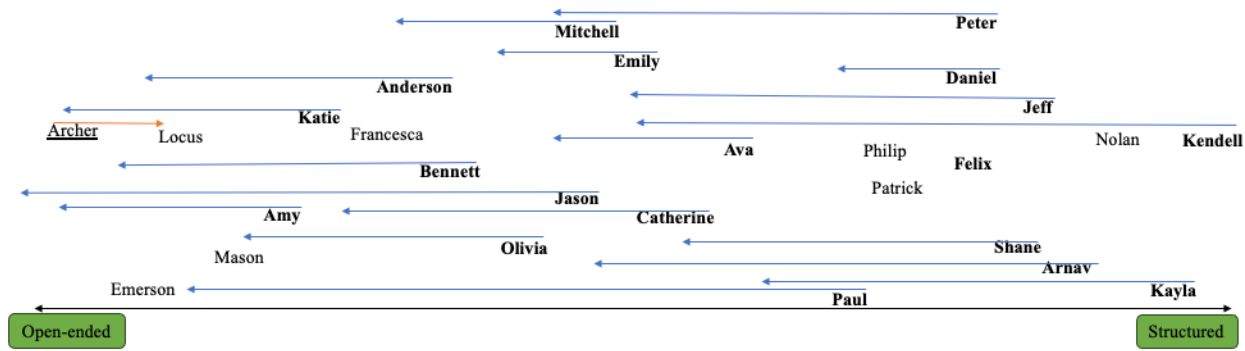

Data from this exercise showcases teachers' inclination to move toward taking a more open-ended approach in teaching labs in the future compared to their current self-assessment of the integration of open-ended labs in their courses. Figure 5 depicts teachers' movement between where they are and where they desire to be in teaching physics labs. As is evident in this figure, the majority i.e., 19 out of 27 teachers moved to the left towards open-ended labs. The magnitude of change was different, though. Pictorial analysis also showed nine teachers did not move from their initial positions, indicating that they felt confident with their lab style in their classes at the time of the rating task. Out of these eight teachers, three teachers were closer to the structural end of the continuum, and five teachers were closer to the open-ended end. One teacher moved to the right- Archer. For instance, Kayla who had moved to the left but still was closer to the right side, explained their positioning the following way she mentioned *"what I have found from nine years of teaching is that different classes have different needs and if you want them to actually get something out of the lab, I have to very clearly structure it in a way that they're going to get something out of that lab. Because I have students that do open-ended labs they are like that was fun and then I'm like what did you learn and they don't tell me anything."*

**RQ3: Evidence of Change from a More Structured Lab to a More Open-ended Approach: A Case Study.** In this phase, we present a compelling case of a teacher whose perceptions and practices showed a change. Dawn is a relatively new teacher with three years of experience in teaching physics in a rural high school. All data collection happened during her second and third year in the PD program (during her first year, she was not teaching physics). Dawn, who holds a degree in biology (out-of-field teacher), initially gravitated toward structured labs driven by a belief about her deficiencies in physics content and technological knowledge particularly with the iOLab device. However, five months after she talked about her deficiencies in the online PD when we approached her to schedule a classroom observation, she surprised us by opting to run an open-ended inquiry lab for the first time in her class. It is worth noting that our routine observations aim not to dictate a particular approach to teaching physics but rather to encourage and support them by providing positive feedback. Dawn's decision to perform an open-ended inquiry lab in the class in front of our cameras was entirely her own. To this end, she adopted a lab that mimics the transit method of observing exoplanets and determining their relative size. In the lab, a lamp has the role of a star, the iOLab's that of a telescope, and beads the role of exoplanets. Students are only asked about the size of the planets (beads), so students do not have to make the beads transit the lamp multiple times. For this lab, students had access to iOLab devices, beads, lamps, and strings in a lab class environment equipped with other items such as measuring tapes and rulers. Below, we present the results of documenting her shift during her journey as an IPaSS teaching fellow.

*First-time Implementation of Open-ended Inquiry Lab: Challenges and Opportunities.* The lab took two class periods which the first period was observed. Dawn provided minimum scaffolding and asked students to come up with a setup to measure the size of different beads that represented exoplanets when they were illuminated by lamps (See Figure 9). This version had been used previously by another fellow (who is not among the participants) at a large, Title 1 suburban school.

The class started by stating the goal by Dawn: determine the size of an unknown exoplanet with the light sensor of iOLab. Then the teacher showed a picture of a possible setup to the students and talked about lab report requirements. Based on the observation notes and video data from observation, it became clear that the lab was guided-inquiry style and not completely open-ended for a few reasons. First, the investigation question was raised by the teacher and not by the student which is a feature of open-ended style labs (Akuma & Collaghan, 2016). The question was: what is the size of an unknown exoplanet? Second, by showing an example of the possible setup and hinting to them that they could use the iOLab light sensor, lamp, beads, and reference "exoplanet", the teacher scaffolded the students' design of the lab. Third, the teacher guided students throughout the lab, although minimally but the students were still seeing her as a source of information. For instance, the teacher hinted that the beads should be installed on a level surface, and they should be moving, just like real exoplanets.

**Figure 9**

*An Example of a Student's Setup for determining the size of the exoplanet with iOLab*

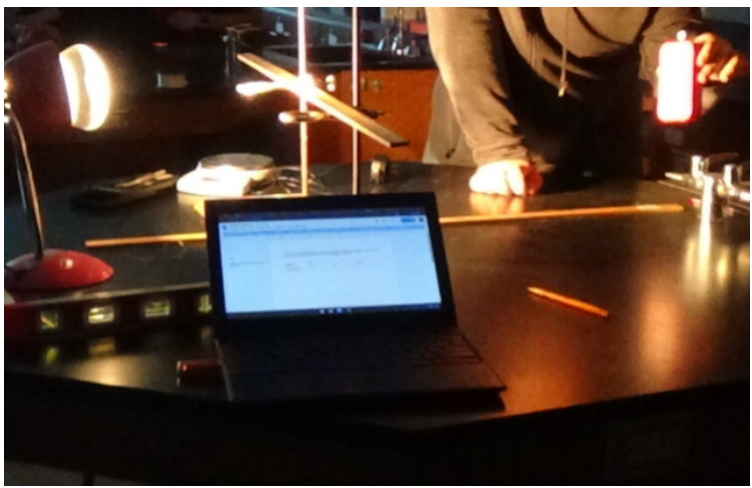

During our post-observation online meeting, I conducted a semi-structured interview to prompt Dawn to talk about her experience and feelings about performing this lab in her class. I started the interview by expressing gratitude and excitement from this visit and then I asked about her overall impression of the lab. Initially, Dawn mentioned that things have gone "pretty good", but she tried to give a real picture by pointing out both positive aspects and challenges. One area of concern she mentioned was with the lab reports that "were a little scary, because I mean it was their first, so." However, she also acknowledged that some students had produced good reports, which was encouraging. Then again she talked about the challenges which she named difficulty in getting students to do the right setup:

> I had quite a few that I kind of had to steer more like they were just having the planet hang in front of the sun. And so I was trying to get them to think it needs to move. Exoplanets move right? So, a few of them I had to push a little bit further on that they didn't quite have that thought process.

In the next prompt, I tried to capture her feelings about this lab by directly asking her how she felt and whether it went based on what she expected or not. In response, she shared that her expectations were higher and she thought the setup was easier for the students:

> I mean it probably went worse than I thought it would. I thought that it would be obvious to how to move the exoplanets across the sun …So I just I don't know. I just kind of assumed that they would all figure that out, especially when I even had the picture of the

> person with the string like this, you know so.... I was a little, I mean I wouldn't say disappointed, but you know I expected more.

It is possible that Dawn was thinking of this lab an as easy one to start but it somehow went against her expectations. Then she talks again about the lab reports, but this time, she also blames herself for providing insufficient scaffolds and not prepping them enough before this lab:

> And then, like I said with the lab reports they were pretty awful just honest right? But I feel like I hadn't given them enough scaffolding to get them to a great lab report, because, like I said, they had hardly written anything in physics, so I think I needed to scaffold it more throughout the year before just throwing it on them.

After hearing Dawn's frustrations with the lab reports and lab setup, I asked whether she would be interested in repeating similar lab styles in her class. She quickly responded to this question with a confidence in her voice:

> Oh, yeah, uhm. I even looked into buying more lamps with my school money this year, so that I can do it with astronomy.

I asked a follow-up question about the gains that make her confident in repeating this style of lab despite challenges. Dawn pointed out a few interesting points here. First, she emphasized the value of observing students' thinking processes which she attributed to the *"openness"* of the task: "I liked watching them think" or "I loved them to be able to think and kind of discuss it with each other. That's always fun." Dawn also reflected that allowing students to figure things out by themselves is a valuable exercise for herself, despite the challenge she faces in resisting the urge to intervene: "*And it is difficult for me very much not to tell them how to do it. (laughing) Like, do it!"*
Second, she talked about the importance of having students do the lab with the iOLab device if they attend U of I:

> And I liked using the iOLab being able to use the iOLab more just because I have so many students that end up going to U of I, maybe not in physics, but they do end up going to the U of I, and if they do end up going into physics at U of I would be nice if they already have the iOLab knowledge, experience. So, I like that.

Third, she brings back the issue of poor lab reports and this time attributes the problem to a bigger issue at the school level where students haven't been adequately prepared across all sciences:

> I think in our school we have not done well at teaching lab reports, and so I think some of our students are going to already start behind in college if they are in a science major. So, I'm hoping to kind of start building that a little bit more into my upper chemistry, like my chemistry and my zoology and astronomy and physics, so that they can have more experience with it.

***Shift in Perception-related Barriers by De-emphasizing One Correct Answer and Disrupting the "Perfect Teacher" Image.*** Five months before Dawn's first implementation of an open-ended inquiry lab, during an online PD (data collected for RQ1 of this study), we opened the floor for teachers to talk about their approaches to teaching physics labs. In that meeting, Dawn talked about her perception-related barriers related to a lack of physics and technology knowledge being on her way of deviating from structured labs:

> It's knowledge-based based and I don't know physics enough to let them just go, because then they ask me questions and I don't know how to answer it. So, it's a comfort level definitely having it laid out especially to like not only I don't know the physics necessarily, I also don't know the iOLab. So, if something goes wrong, I don't understand why it's showing something.

During the interview, when I reminded her of her early concerns, she still repeated the same concerns adding "I think it's always daunting." However, this time she talked about two approaches that made her confident to take a risk and perform an open-ended inquiry lab in her class. First, she talked about the importance of de-emphasizing the right answer by taking an iterative approach to taking data, improving the experiment, and repeating the data collection:

> Doing this lab specifically, it really helped like maybe they didn't get the right answer. And I kept telling them, you know, it's okay if you don't get the right answer as long as we get the actual data, and we don't fudge our data, right? (laugh) And we get the conclusions from the data. And we learn to improve the experiment. And so, I think that just even having the experience like it gave me a little bit more confidence in being able to do that,

Then, she talked about the importance of collecting incorrect data as an entry point which enabled her to make a point about different setups:

> I didn't know if the hanging down planet not moving would show the same results as a moving planet did. and so just letting them do it and let's see, you know. And it was okay, and it worked, and so and but I could show them. You know that your data wasn't as accurate as this one's data because his was moving. You know so.

The second approach that boosted Dawn's confidence involved disrupting the notion of "perfect teacher". Seeing more experienced teachers in PD sessions who are still in their learning journeys shifted Dawn's perception away from waiting to accumulate more experience to taking a more open-ended approach:

> I mean definitely… Seeing that even extremely experienced teachers don't have it all and don't know it all. I would think that all of the teachers in this PD would tell you they don't know it all, and they don't teach perfectly, and they are…. I would say, even most of us probably wouldn't even say that we're good teachers which to say like we do what we can. Like Arnav, you know, saying that he has so much to learn, and he is so close to retirement. And it's like realizing that the perfect teacher that I've built up in my head does not exist, and you don't have to be perfect to start. You know that we are still learning. Yeah, every year, even our last year, before we retire, you know. Carl was retired. So, you know, he was learning up until so.

Being a teaching fellow in the IPaSS CoP and witnessing that even experienced teachers sometimes encounter challenges, changed Dawn's perceptions about herself and her capabilities. Seeing the experience and knowledge gap diminish, Dawn now thinks she is nothing less than other teachers, and if they could take a more open-ended approach, so could she.

Along the same lines, in open-ended survey results (collected for RQ1), Dawn also emphasized the importance of engaging with IPaSS teachers in the community:

It [the community] has helped me immeasurably.  First and foremost, helping me understand the physics better. Second, hearing all the different ways of the approach to teaching. I feel like I have learned how to step back more (still need a lot of work) and let them figure things out.

***Sharing the Experience with the Community.*** Before the summer PD, IPaSS teachers are encouraged to present something from the past school year at the PD. To help teachers with this process and ensure the summer PD program is coherent, PD facilitators and teachers have Zoom meetings in spring to plan teachers' presentations. The thirteen-minute planning meeting with Dawn was recorded, partially transcribed, and analyzed. Dawn chose to present her exoplanet lab to her colleagues. One thing she mentioned in the meeting was how giving students the liberty to come up with a design gave rise to many different ways of doing the lab: "I wish I had taken more pictures when the students were doing it because some of them came up with some different ways of doing it." In this meeting, it was determined Dawn would share the student brainstorming phase with her IPaSS peers, so they could get a sense of how she starts the lab.

The IPaSS CoP convinced Dawn that it is okay not to be "the perfect teacher" and thereby it gave Dawn confidence that she was capable of asking her students to do an open-ended lab. Dawn's successful implementation of the lab in her class with the observation of *students' different ways of doing it* allowed her to showcase it as a successful example of open-ended labs to other physics teachers in the community. In the in-person PD session, Dawn introduced her adapted lab and talked about this experience as a successful example of an open-ended lab illustrating students' scientific thinking. This case shows how profound the influence of a CoP can be on teachers like Dawn, facilitating transformative shifts in perceptions that are reflected in classroom practices.

## Discussion

In this work, we investigated the barriers to taking a more open-ended approach in teaching labs among high school physics teachers after removing some prevalent barriers such as access to lab equipment and a community for support. We further examined the role of the community in instigating change among teachers and documented a case as evidence of overcoming seemingly persistent barriers for a novice teacher.

**Barriers to Taking an Open-ended Approach in Teaching Labs in Physics Classes**

The findings of this study revealed that, despite having access to a Community of Practice (CoP) for support and the iOLab to facilitate an open-ended inquiry approach, teachers continued to face structural and perception-related barriers when trying to shift from their current structured lab format to a more open-ended approach. These barriers were consistent with those identified in previous research, where teachers did not benefit from social and material support (Anderson, 2007; Deters, 2004; Ramnarain & Schuster, 2014). Regardless of their experience or expertise levels, teachers reported facing structural barriers, perception-related barriers, or a combination of both after participating in the program for at least one year (100 hours). The results also indicated that perception-related barriers concerning teachers' views of their own physics content and technological knowledge were primarily found among novice teachers with non-physics backgrounds. Experienced teachers and those with physics backgrounds found structural barriers, such as short class periods, to be more problematic.

While a common approach might involve simply teaching novice teachers specific physics content, this method may not fully instill the desired level of confidence in their teaching abilities. Findings from our case study suggest that teachers can build confidence in CoP even without mastering every aspect of the content. By observing more experienced teachers and faculty in physics who are still learning and acknowledging that learning is a collaborative process where people with different levels of knowledge can grow together, novice teachers can feel more assured in their abilities. For example, Dawn no longer waits to achieve a certain level of knowledge or experience before adopting a more open-ended approach. Instead, she embraces the risk, understanding that she is not alone, and does not attribute any lack of success to her background.

Another important finding here was that when perception-related barriers are addressed, structural barriers are subsequently resolved. Therefore, perception-related barriers hold greater significance. In our case study of Dawn, once she overcame her content and technological barriers, she no longer reported a lack of time in class for taking an open-ended approach.

**Impact of Professional Development and Support on Changing Perception and Practices**

Different data sources in this study revealed that support from the community has indeed helped teachers in changing their perceptions and overcoming barriers. The retrospective survey results, embodied rating task, and case study provide evidence that there is a gap between teachers' desired and current positions in teaching labs in terms of taking an open-ended approach. However, challenges exist for teachers hindering them from progressing in the direction they desire. However, support mechanisms work in particular ways to help teachers overcome implementation barriers of open-ended labs in their classes.

By investigating the role of the CoP in this work, we aimed to understand how teachers who "desire" to move toward a more open-ended approach, get help from the community, overcome barriers, and change their practices. In this regard, our results highlighted two points. First, we learned that showing vulnerability by experienced teachers disrupted the view of the "perfect teacher" for novice teachers like Dawn. The case study results revealed that just being in the community and interacting with more experienced teachers *do not* guarantee change. What instigated change for our novice teacher, was witnessing the learning journeys of veteran members with all the challenges and failures they still face. Disrupting the image of a "perfect teacher" for Dawn was an inflection point where she found herself *enough* to take risks in her class. The specific examples of Arnav and Carl that Dawn mentioned here are the stories of sharing vulnerability by more experienced teachers, too that helped in the same way. Previously, our work with the same population of teachers had shown the importance of showing vulnerability by veteran teachers in opening communication doors toward better learning and support (Mahmood et al., 2024).

Second, we learned that teachers need time to develop trust with the CoP for the change to happen at the perception and practice levels. Some teachers need a significant amount of time to feel comfortable making sustainable changes to their practices. For Dawn, this duration was as long as two years. Depending on their background and the structure of the PD and community activities, this duration may vary. Hence, the benefits of community involvement do not arise immediately after joining or by membership in a community per se. This work adds to the literature in favor of prolonged PD for in-service and pre-service teachers (Borko, Jacobs, & Koellner, 2010) by emphasizing the importance of prolonged PD activities in a responsive way for novice teachers and teachers with diverse science backgrounds.

**Considerations for Designing Professional Development**

Several studies in the literature underscore the importance of professional development for supporting reform-based teaching practices by using strategies such as weekly meetings, pre-semester workshops, and building a community of learners (Friedrichsen et al. 2006; Hofstein et al. 2004; Lazarowitz & Tamir, 1994; Rutberg, 2023). While echoing these recognized strategies and designing them into a prolonged PD, we argue that giving teachers epistemic agency (Sundstrom, Phillips, & Holmes, 2020; Smith, Stein, & Holmes, 2018) -cognitive authority to decide which knowledge is valuable (Sundstrom et al. 2023) in their context - by flexible implementation of materials is key. In this approach, which we call responsive professional development elsewhere, we encourage professional developers to attend to teachers' needs and design PD experiences based on those needs. One manifestation of attentiveness, highlighted in Dawn's narrative, involved the flexible implementation of materials without requiring her to follow a prescribed timeline. This flexibility offered Dawn an absorption period that lasted for two years before she decided to implement an open-ended approach in teaching a physics lab. We surmise that this flexible approach from PD facilitators may be a particularly important component in supporting novice teachers and those with non-physics backgrounds. Future works should consider PD structures such as flexibility in studying teacher perception and practice change.

To support teachers in removing some structural barriers, it is important to have them take a more open-ended approach in doing labs during PD sessions. Considering the importance of this strategy, the IPaSS program, created designated time and space during in-person PD sessions for teachers to try new ways of doing labs before testing in their class. We suggest PD programs encourage teachers to work through the labs in as many different ways as they can consider prior to implementation and build space to do this in the PD. Additionally, programs could benefit from creating a repository for each lab where teachers can document pedagogical, technological, and physics content challenges that arose during implementation so that teachers implementing that lab in the future will begin with a solid baseline.

### Conclusions and Implications

The findings of this study contribute to understanding the complexities of implementing new pedagogical practices in science education and open-ended labs in physics education. By presenting data representing collective changes in teachers' perceptions of open-ended labs, paired with qualitative evidence of how teachers overcome barriers, this study shows the power of the community in catalyzing pedagogical change. Additionally, the narrative of individual changes within a program that advocates for flexibility resonates with the broader narrative of recognizing teachers as capable agents who can gradually evolve when given liberty in choice. In the end, although the effectiveness of investigative science learning approaches has been proven effective in many works, it is important to consider the desired outcomes of the course, and students' and teachers' content and technological knowledge when it comes to taking a more open-ended approach in teaching labs.

### References [will be inserted]